\documentclass[aps,pra,twocolumn,nopacs,letterpaper]{revtex4}

\usepackage{amsmath}
\usepackage{color,graphicx}
\usepackage[pdfstartview=FitH,colorlinks=true,linkcolor=blue,citecolor=blue,urlcolor=blue]{hyperref}

\renewcommand{\vec}[1]{\bf #1 \rm}
\newcommand{\q}{\vec{q}}
\newcommand{\up}{\uparrow}
\newcommand{\down}{\downarrow}

\begin{document}

\title{Spin-Imbalanced Fermi Superfluidity in a Lieb Lattice Hubbard Model}

\author{Marek Tylutki}
\email{marek.tylutki@aalto.fi}

\author{P\"{a}ivi T\"{o}rm\"{a}}
\email{paivi.torma@aalto.fi}

\affiliation{Department of Applied Physics, Aalto University, Helsinki, Finland}

\date{\today}

\begin{abstract}
We obtain a phase diagram of the spin imbalanced Hubbard model on the Lieb lattice, which is known to feature a flat band in its single-particle spectrum. Using the BCS mean-field theory for multiband systems, we find a variety of superfluid phases with imbalance. In particular, we find four different types of FFLO phases, i.e. superfluid phases with periodic spatial modulation. They differ by the magnitude and direction of the centre-of-mass momentum of Cooper pairs. We also see a large region of stable Sarma phase, where the density imbalance is associated with zero Cooper pair momentum. In the mechanism responsible for the formation of those phases, the crucial role is played by the flat band, wherein particles can readjust their density at zero energy cost. The multiorbital structure of the unit cell is found to stabilize the Sarma phase by allowing for a modulation of the order parameter within a unit cell. We also study the effect of finite temperature and a lattice with staggered hopping parameters on the behaviour of these phases. 
\end{abstract}

\maketitle

\section{Introduction}
Electric and thermal conductivity in metals, as well as the phenomenon of superconductivity are typically modelled by motion of electrons in the field of a static crystal lattice, which is often square or cubic (depending on dimensionality), or there is no lattice at all. This simple picture was sufficient for understanding, for example, low temperature superconductivity. To this end, a mean-field theory was devised, named after the authors the Bardeen-Cooper-Schrieffer theory (BCS)~\cite{Bardeen1957a}. In the conventional superconductor with interactions through $s$-wave scattering, the emerging two-body correlations can be pictured as pairs of electrons with opposite spins and momenta (Cooper pairs)~\cite{Cooper1956} that develop an off-diagonal long-range order at low temperatures. This pairing mechanism and its developments lead to multiple new phases of matter, such as conventional and unconventional superconductivity~\cite{Bednorz1986,Lee2006,Bloch2008}, including topological superconductors~\cite{Sato2017}. The same pairing mechanism explains unconventional superfluid phases in liquid $^3$He~\cite{Leggett1975}. Apart from Cooper pair formation, interacting fermions may also be subject to Fermi surface (FS) instability towards symmetry-breaking deformations of the FS, called the Pomeranchuk instability~\cite{Pomeranchuk1959}. 

Going beyond the standard square lattice in modelling condensed matter phenomena adds to this variety of non-trivial phases, in particular to exotic superfluidity. Among such non-trivial superfluid phases the Fulde-Ferrell-Larkin-Ovchinnikov (FFLO) family of phases stands out, where the pairing between two imbalanced spin components is possible due to the Cooper pairs acquiring non-zero centre-of-mass momentum~\cite{Fulde1964,Larkin1965,Kinnunen2017}; the mechanism can be pictured as a relative momentum shift of the non-interacting Fermi surfaces of the two spin components. Such a scenario was proposed by Fulde and Ferrell (FF)~\cite{Fulde1964} and independently by Larkin and Ovchinnikov (LO)~\cite{Larkin1965} for systems in spatial continuum. In the first ansatz, Cooper pairs have a single momentum, ${\bf q}$, while in the second both ${\bf q}$ and $-{\bf q}$ are possible. Both versions lead to a spatial, periodic modulation of the order parameter, but the LO ansatz is also characterized by a spatial variation of density. This continuum model can, in principle, be realized with ultracold Fermi gases, where the scattering length is appropriately tuned. However, those predictions have been rather elusive, supported only by indirect experimental evidence~\cite{Casalbuoni2004,Liao2010}. In ultracold quantum gases, phase separation has been observed instead of exotic spin-imbalanced superfluids~\cite{Zwierlein2006s,Partridge2006,Partridge2006a,Shin2008,Nascimbene2009}, consistently with predictions for continuum systems~\cite{Machida2006,Sheehy2006,Sheehy2007,Recati2008}. As an alternative, a deformed FS superfludity~\cite{Muether2002,Sedrakian2005} has been proposed as giving a lower energy than the conventional BCS theory.

The paradigmatic model of fermions in a lattice potential is the Hubbard model~\cite{Torma2015}, where fermions can interact only on-site. In analogy to continuum systems, the attractive Hubbard model supports the Cooper pair formation and the mean-field BCS theory succeeds in explaining superfluidity~\cite{Torma2015}, including models with spin-orbit coupling~\cite{Iskin2013,Qu2013}. Since the FS of a Fermi gas in a square or cubic lattice loses its isotropy and has a square-like shape in two dimensions, especially close to Van Hove singularities, it facilitates the FFLO pairing due to a nesting effect~\cite{Kinnunen2017}, in which the pairing takes place where the FSs, shifted by the momentum $\q$, match. Therefore, relatively significant regions with non-uniform superfluidity have been found theoretically, both with the FF~\cite{Koponen2006,Koponen2007,Wolak2012,Cichy2018,Heikkinen2014} and the LO~\cite{Baarsma2016} ansatz. In the repulsive Hubbard model, superfluidity may coexist with the magnetic stripe order as found by iPEPS~\cite{Corboz2014} and DMFT methods~\cite{Vanhala2017}, or with Pomeranchuk instability as in Refs.~\cite{Kiesel2013,Kitatani2017}. 

Another way in which the spin imbalance can be accommodated into superfluidity is the Sarma mechanism. It was first proposed by Sarma as an unstable solution to the gap equation~\cite{Sarma1963} and then reintroduced by Liu and Wilczek at fixed chemical potentials, wherein the pairing takes place between atoms on the FS of the minority component and the atoms inside the Fermi sea of the majority component; therefore, it is also known as the interior gap pairing~\cite{Liu2003}. However, at fixed chemical potentials, this phase turns out to be unstable; this phase was theoretically found to be unstable unless fully gapped in continuum~\cite{Wu2003,Sheehy2006}. Only later it was realized~\cite{Forbes2005,Pao2006,Gubbels2006} that for the stability of the Sarma phase fixed densities or an external trapping potential are required. The existence of a stable gapless superfluid was postulated also in Ref.~\cite{Son2006} via an effective theory in the vicinity of the phase diagram splitting point. It was also found stable in a general two-band model~\cite{He2009}, in a lattice~\cite{Koponen2006,Koponen2007} at constant densities and recently in a mixed geometry lattice in Ref.\cite{Kim2013}. A method for the detection of the Sarma phase was proposed in Ref.~\cite{Yi2006}. 

Including orbital degrees of freedom into a lattice, which now has more than just one band in its spectrum, increases the number of ways in which the pairing can take place~\cite{Kim2013}. Among these multiband systems particularly interesting are lattices that feature a flat band (FB) in their single-particle spectrum~\cite{Leykam2018,Peotta2015,Julku2016}. Such a FB allows the atoms residing therein to change their momentum distribution freely, i.e. with no energy cost. Thus, a large variety of configurations is accessible for those atoms, and they may minimize their energy more easily by developing pairing correlations. FBs are known to enhance superfluidity, and they can also facilitate non-uniform superfluid phases due to the interband pairing, as recently reported in Ref.~\cite{Huhtinen2018}. In this article, we consider one such flat-band lattice, a Lieb lattice, which comprises three sites in its unit cell, and we further develop the ideas introduced in Ref.~\cite{Huhtinen2018}. Such lattices have been experimentally realized for ultracold gases~\cite{Taiee1500854,PhysRevLett.118.175301}, in designer lattices~\cite{Slot2017,Drost2017}, and localized flat band states have been observed in optical analogues~\cite{PhysRevLett.114.245504,PhysRevLett.114.245503}.

In Section~\ref{sec.model} we present the studied model and the method we use, in Sec.~\ref{sec.phd} we calculate the phase diagram for symmetric hoppings, and we identify a plethora of non-trivial superfluid phases with spin imbalance. Thereafter, we focus on mechanisms behind those phases: FF and $\eta$ phases in Sec.~\ref{sec.ff} and Sarma phase in Sec.~\ref{sec.sarma}. Finally, we show the effect of temperature in Sec.~\ref{sec.temp} and the effect of staggered, non-symmetric hopping in Sec.~\ref{sec.stag}. We summarize our results in Sec.~\ref{sec.conclusions}. 

\begin{figure}[!h]
\includegraphics[width=.95\columnwidth]{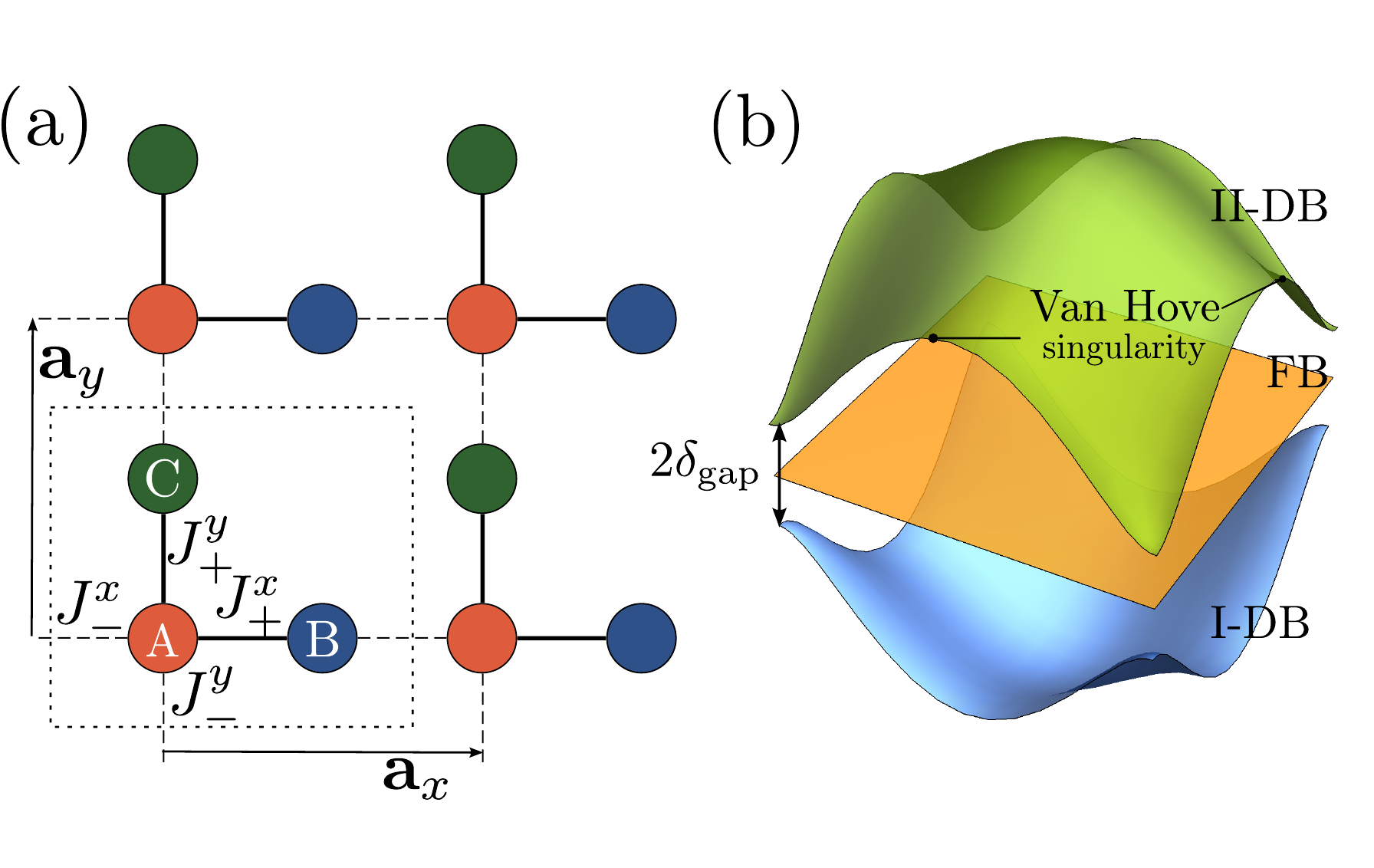}
\caption{(a) The Lieb lattice has three sites in the unit cell, which may assume different values of density or order parameter for spin-imbalanced superfluid phases. In Sec.~\ref{sec.stag} we consider hoppings $J^{x, y}_\pm$ that may have different values in different directions. (b) The band structure of the lattice: There are three bands, two dispersive bands and a flat band in the centre. This band structure is symmetric with respect to the zero energy plane, where the flat band resides. The FB and the Van Hove singularities give rise to a diverging density of states. If the hoppings $J^{x, y}_\pm$ are not all equal, a gap $\delta_{\rm gap}$ opens thus isolating the FB. }
\label{fig.schematic}
\end{figure}
\section{Model}
\label{sec.model}
We study a spin-imbalanced Fermi gas on the Lieb lattice, which is described by the Hubbard model, 
\begin{equation}
H = \sum_\sigma \sum_{\bf i \alpha, j \beta} \psi_{\bf i \alpha \sigma}^\dag \mathcal{H}_{\bf i \alpha, j \beta}^{\vphantom{\dag}} \psi_{\bf j \beta \sigma}^{\vphantom{\dag}} - \sum_{\sigma} \mu_\sigma N_\sigma + H_{\rm int} ~,
\label{eq.hamilt}
\end{equation}
where the kinetic term, $\mathcal{H}_{\bf i \alpha, j\beta}$, describes the hopping between neighbouring sites, the on-site interaction term reads as
\begin{equation}
H_{\rm int} = U \sum_{\bf i \alpha} \psi_{\bf i \alpha \up}^\dag \psi_{\bf i \alpha \down}^\dag \psi_{\bf i \alpha \down}^{\vphantom{\dag}} \psi_{\bf i \alpha \up}^{\vphantom{\dag}} ~,
\end{equation}
and the particle number is given by $N_\sigma = \sum_{\bf i \alpha} \psi_{\bf i \alpha \sigma}^\dag \psi_{\bf i \alpha \sigma}^{\vphantom{\dag}}$. The Lieb lattice, shown in Fig.~\ref{fig.schematic}(a), has three sites in its unit cell (which we label as A, B and C), that is spanned by the lattice vectors ${\bf a}_x$ and ${\bf a}_y$. The latin indices, e.g. $\bf i$, are coordinates of the sites in space, while the greek indices, $\alpha$, number the sublattices: A, B or C. We include chemical potentials, as we consider the grand-canonical ensemble. We define the average chemical potential as $\mu = (\mu_\up + \mu_\down) / 2$ and the effective magnetic field as $h = (\mu_\up - \mu_\down) / 2$. We also allow for a dimerization of the bonds of the lattice in both directions, which means alternating strong and weak hoppings as shown in Fig.~\ref{fig.schematic}(a),  
\begin{equation}
J^{x, y}_\pm = (1 \pm \delta_{x, y}) J ~.
\end{equation}
The parameters $\delta_{x, y}$ are called hopping staggering parameters. We use the value of $J$ as the unit of energy and the magnitude of lattice vectors $a = |{\bf a}_x| = |{\bf a}_y|$ as the unit of length wherever units are not explicitly stated. 

The spectrum of the non-interacting system, at $U = 0$, is determined by a single-particle Hamiltonian. By performing a Fourier transform
\begin{equation}
\psi_{\bf j \alpha \sigma} = \frac1{\sqrt{V}} \sum_{\bf k} c_{\bf k \alpha \sigma} e^{i \bf k j} ~,
\label{eq.ft}
\end{equation}
where $\sigma = \up, \down$; ${\bf k} = (k_x, k_y)$ and $V$ is the system's volume, and by collecting the fields as $\psi_{\bf k} = (c_{{\bf k}, A \up}^{\vphantom{\dag}}, c_{{\bf k}, B \up}^{\vphantom{\dag}}, c_{{\bf k}, C \up}^{\vphantom{\dag}})^T$, the non-interacting Hamiltonian has a compact from $H = \sum_{\bf k} \psi^\dag_{\bf k} \mathcal{H}^{\phantom{\dag}}_{\bf k} \psi^{\phantom{\dag}}_{\bf k}$, 
where
\begin{equation}
\mathcal{H}_{\bf k} = -2 J \begin{bmatrix}
0 & s & t \\
s^* & 0 & 0 \\
t^* & 0 & 0 \\
\end{bmatrix} ~,
\end{equation}
with $s = \cos \frac{k_x}{2} + i \delta_x \sin \frac{k_x}{2}$ and $t = \cos \frac{k_y}{2} + i \delta_y \sin \frac{k_y}{2}$. The non-interacting Hamiltonian gives the band structure: the flat band (FB), $\varepsilon({\bf k}) =  0$, and two dispersive bands $\varepsilon({\bf k}) =$
\begin{equation*}
\pm J\sqrt2 \sqrt{2 + (1 - \delta^2_x) \cos k_x a + (1 - \delta^2_y) \cos k_y a + \delta^2_x + \delta^2_y} ~, 
\end{equation*}
hereafter denoted as I-DB for the lower and II-DB for the upper one. This band structure gives rise to two types of singularities in the density of states, one corresponding to a flat band and the other to a diverging density of states in either of the dispersive bands (Van Hove singularities at $\varepsilon = \pm 2\, J$). When either of $\delta_{x,y}$ is non-zero, there is a band gap in the single-particle energy spectrum on both sides of the FB, whose width is given by
\begin{equation}
\delta_{\rm gap} = 2J\, \sqrt{\delta^2_x + \delta^2_y} ~.
\end{equation}
When the hopping dimerization parameters are set to zero, $\delta_{x, y} = 0$, the gap closes. The first Brillouin zone has three types of high symmetry points: a $\Gamma$ point at ${\bf k} = (0, 0)$, an $M$ point at ${\bf k} = (\pm \pi, \pm \pi)$ and two types of $X$ points at ${\bf k} = (0, \pm \pi)$ or $(\pm \pi, 0)$. The band structure is shown in Fig.~\ref{fig.schematic}(b). 

We treat the interaction with a mean-field method, introducing a BCS pairing field $\Delta_{\bf i \alpha} = U \langle \psi_{\bf i \alpha \down} \psi_{\bf i \alpha \up} \rangle$. The system is at finite temperature, and we calculate the averages over the grand-canonical ensemble, i.e. $\langle \mathcal{O} \rangle = {\rm Tr}\, [\mathcal{O} e^{-\beta H}] / {\rm Tr}\, e^{-\beta H}$, with $\beta = (kT)^{-1}$ the inverse temperature and $k$ the Boltzmann constant. With these definitions, the interaction becomes a quadratic function of the fermionic fields,
\begin{equation}
H_{\rm int} = \sum_{\bf i \alpha} \Delta_{\bf i \alpha}^{\phantom{\dag}} \psi_{\bf i \alpha \up}^\dag \psi_{\bf i \alpha \down}^\dag + {\rm H.c.} - \frac1{U} \sum_{\bf i \alpha} |\Delta_{\bf i \alpha}|^2 ~.
\label{eq.hint}
\end{equation}
We make the Fulde-Ferrell ansatz for the order parameters, $\Delta_{\bf i \alpha} = \Delta_\alpha e^{i\, \bf q\, i}$, where the vector $\q$ plays a role of the centre-of-mass momentum of Cooper pairs (the FF momentum). Upon performing a Fourier transform~(\ref{eq.ft}) and collecting the fields into a Nambu spinor,
\begin{equation*}
\Psi_{\bf k} = (c_{{\bf k}, A \up}^{\vphantom{\dag}}, c_{{\bf k}, B \up}^{\vphantom{\dag}}, c_{{\bf k}, C \up}^{\vphantom{\dag}},
c_{{\bf q - k}, A \down}^\dag, c_{{\bf q - k}, B \down}^\dag, c_{{\bf q - k}, C \down}^\dag)^T ~,
\end{equation*}
the mean-field Hamiltonian becomes
\begin{equation}
H_{\rm FF} = \sum_{\bf k}\, \left[  \Psi^\dag_{\bf k} \mathcal{H}_{\rm BdG}^{\vphantom{\dag}} \Psi_{\bf k}^{\vphantom{\dag}} - 3 \mu_\downarrow -  \frac1U {\rm Tr}\, \boldsymbol{\Delta}^\dag \boldsymbol{\Delta} \right] ~,
\end{equation}
with the Bogoliubov-de Gennes (BdG) Hamiltonian conveniently defined as
\begin{equation}
\mathcal{H}_{\rm BdG} = \begin{bmatrix}
\mathcal{H}_{\bf k} - \mu_{\uparrow} & \boldsymbol{\Delta}\\
\boldsymbol{\Delta}^\dag & -\mathcal{H}_{\bf -k+q} + \mu_{\downarrow}\\
\end{bmatrix} ~.
\end{equation}
The order parameters are collected into a matrix as $(\boldsymbol{\Delta})_{\alpha \beta} = \delta_{\alpha \beta} \Delta_\alpha$ (no summation). 

Next, we can transform the fields to the non-interacting band basis, defined as $\mathcal{G}^\dag_{\bf k \sigma} \mathcal{H}_{\bf k \sigma} \mathcal{G}_{\bf k \sigma} = \boldsymbol{\varepsilon}_{\bf k \sigma}$, which gives
\begin{equation}
\begin{bmatrix} {\bf d}_{\bf k \up}^{\vphantom{\dag}}\\ {\bf d}_{\bf q - k \down}^\dag\\ \end{bmatrix} = \begin{bmatrix}
 \mathcal{G}_{\bf k \up}^\dag & 0\\
0 &  \mathcal{G}_{\bf q - k \down}^\dag\\
\end{bmatrix} \Psi_{\bf k} ~.
\end{equation}
The operators $({\bf d}_{\bf k \up}^{\vphantom{\dag}}, {\bf d}_{\bf q - k \down}^\dag)^T$ correspond to particles in different bands. Now we perform a further unitary transformation to a quasi-particle basis,
$({\bf \gamma}_{\bf k, q \up}^{\vphantom{\dag}},{\bf \gamma}_{\bf k, q \down}^\dag)^T$,
\begin{equation*}
\begin{bmatrix}
{\bf \gamma}_{\bf k, q \up}^{\vphantom{\dag}}\\
{\bf \gamma}_{\bf k, q \down}^\dag \\
\end{bmatrix} = \begin{bmatrix}
U_{\bf k, q} & -V^\dag_{\bf k, q} \\
V_{\bf k, q} & U^\dag_{\bf k, q} \\
\end{bmatrix} \begin{bmatrix}
{\bf d}_{\bf k \up}^{\vphantom{\dag}}\\
{\bf d}_{\bf q - k \down}^\dag\\
\end{bmatrix} ~,
\end{equation*}
which diagonalizes the full BdG Hamiltonian, $\mathcal{H}_{\rm BdG}$. The transformation matrix retains its particle-hole symmetric structure, as the imbalance, $h$, commutes with the BdG Hamiltonian, $\mathcal{H}_{\rm BdG}$. The diagonalized Hamiltonian reads
\begin{equation}
H_{\rm FF} = \sum_{\bf k} (\gamma_{\bf k, q \up}^\dag {\bf E}_{\bf k,
  q \up}^{\vphantom{\dag}} \gamma_{\bf k, q \up}^{\vphantom{\dag}} +
\gamma_{\bf k, q \down}^\dag {\bf E}_{\bf k, q
  \down}^{\vphantom{\dag}} \gamma_{\bf k, q \down}^{\vphantom{\dag}})
+ \mathcal{E} ~,
\end{equation}
where ${\bf E}_{\bf k, q\, \sigma}$ are numerically calculated diagonal matrices of the quasi-particle energies, and the energy offset reads as
\begin{equation}
\mathcal{E} = -\sum_{\bf k} (3 \mu_\downarrow + \frac1U {\rm Tr}\, \boldsymbol{\Delta}^\dag \boldsymbol{\Delta} + {\rm Tr}\, {\bf E}_{\bf k, q \down}) ~.
\end{equation}

We look for the global minima of the thermodynamic potential, 
\begin{equation*}
\Omega = -\frac1{V \beta} \ln \mathrm{Tr}\, e^{-\beta H_{\rm FF}} ~,
\end{equation*}
which determine the stable phases of the system. In the BCS theory, the thermodynamic potential can be calculated as
\begin{equation}
\Omega = -\frac{1}{V \beta} \sum_{\bf{k}, \sigma}{\rm Tr} \ln \left( 1 + e^{-\beta\, {\bf E}_{\bf k, q\sigma}} \right) + \frac{\mathcal{E}}{V} ~.
\label{eq.thermpot}
\end{equation}
We independently minimize it with respect to all components of $\boldsymbol{\Delta}$ and $\bf q$.

Since the FF ansatz we use limits the possible momentum of Cooper pairs to only one value, $\q$, we return to the original formulation in real space in order to also minimize the original Hamiltonian in Eq.~(\ref{eq.hamilt}) with the mean-field interaction $H_{\rm int}$ given by~(\ref{eq.hint}). This corresponds to including an arbitrary number of possible Cooper pair momenta. To this end, we minimize the energy,
\begin{equation*}
H = \sum_{\bf i \alpha, j \beta} \Psi^\dag_{\bf i \alpha} \begin{bmatrix}
\mathcal{H}_{\bf i \alpha, j \beta} - \mu_\up & \boldsymbol{\Delta}_{\bf i \alpha, j \beta}\\
\boldsymbol{\Delta}^\dag_{\bf i \alpha, j \beta} & -\mathcal{H}_{\bf i \alpha, j \beta} + \mu_\down\\
\end{bmatrix} \Psi_{\bf j \beta} + \mathcal{E}_r ~,
\end{equation*}
with the space dependent spinor $\Psi_{\bf i \alpha} = (\psi_{\bf i \alpha \up}, \psi^\dag_{\bf i \alpha \down})^T$, and $\mathcal{E}_r = -\sum_{\bf i \alpha} |\Delta_{\bf i \alpha}|^2 / U - V \mu_\down$, that at $T = 0$, can be calculated as
\begin{equation*}
\langle E \rangle = \sum_{E_\eta < 0} E_\eta - \frac1U \sum_{\bf i \alpha} |\Delta_{\bf i \alpha}|^2 - V \mu_\down.
\end{equation*}
We diagonalize the Bogoliubov de-Gennes Hamiltonian in real space, 
\begin{subequations}
\label{eq.realbdg}
\begin{equation}
\sum_{\bf j \beta} \begin{bmatrix}
\mathcal{H}_{\bf i \alpha, j \beta} - \mu_\up - E_\eta & \boldsymbol{\Delta}_{\bf i \alpha, j \beta}\\
\boldsymbol{\Delta}^\dag_{\bf i \alpha, j \beta} & -\mathcal{H}_{\bf i \alpha, j \beta} + \mu_\down - E_\eta\\
\end{bmatrix} \begin{bmatrix}
u_{\eta;\, \bf j \beta} \\ v_{\eta;\, \bf j \beta}\\
\end{bmatrix} = 0 ~,
\end{equation}
where the order parameter is now collected into a matrix as $\boldsymbol{\Delta}_{\bf i \alpha, j \beta} = \delta_{\bf i \bf j} \delta_{\alpha \beta} \Delta_{\bf i \alpha}$ (no summation), self-consistently with the gap equation, 
\begin{equation}
\Delta_{\bf i \alpha} = U \sum_{\eta} f(E_\eta, \beta)\, u_{\eta;\, \bf i \alpha}\, v^*_{\eta;\, \bf i \alpha} ~.
\end{equation}
\end{subequations}
The function $f(\epsilon, \beta) = (1 + \exp(\beta \epsilon))^{-1}$ is the Fermi-Dirac distribution function. The densities are calculated as
\begin{align*}
n_{\up \bf i \alpha} &= \sum_{\eta} f(E_\eta, \beta)\, |u_{\eta;\, \bf i \alpha}|^2 ~,\\
n_{\down \bf i \alpha} &= \sum_{\eta} f(-E_\eta, \beta)\, |v_{\eta;\, \bf i \alpha}|^2 ~.\\
\end{align*}
The real space solution of the mean-field equations should converge to the results calculated in momentum space as the lattice size becomes large. 

\begin{figure}[!h]
\includegraphics[width = \columnwidth]{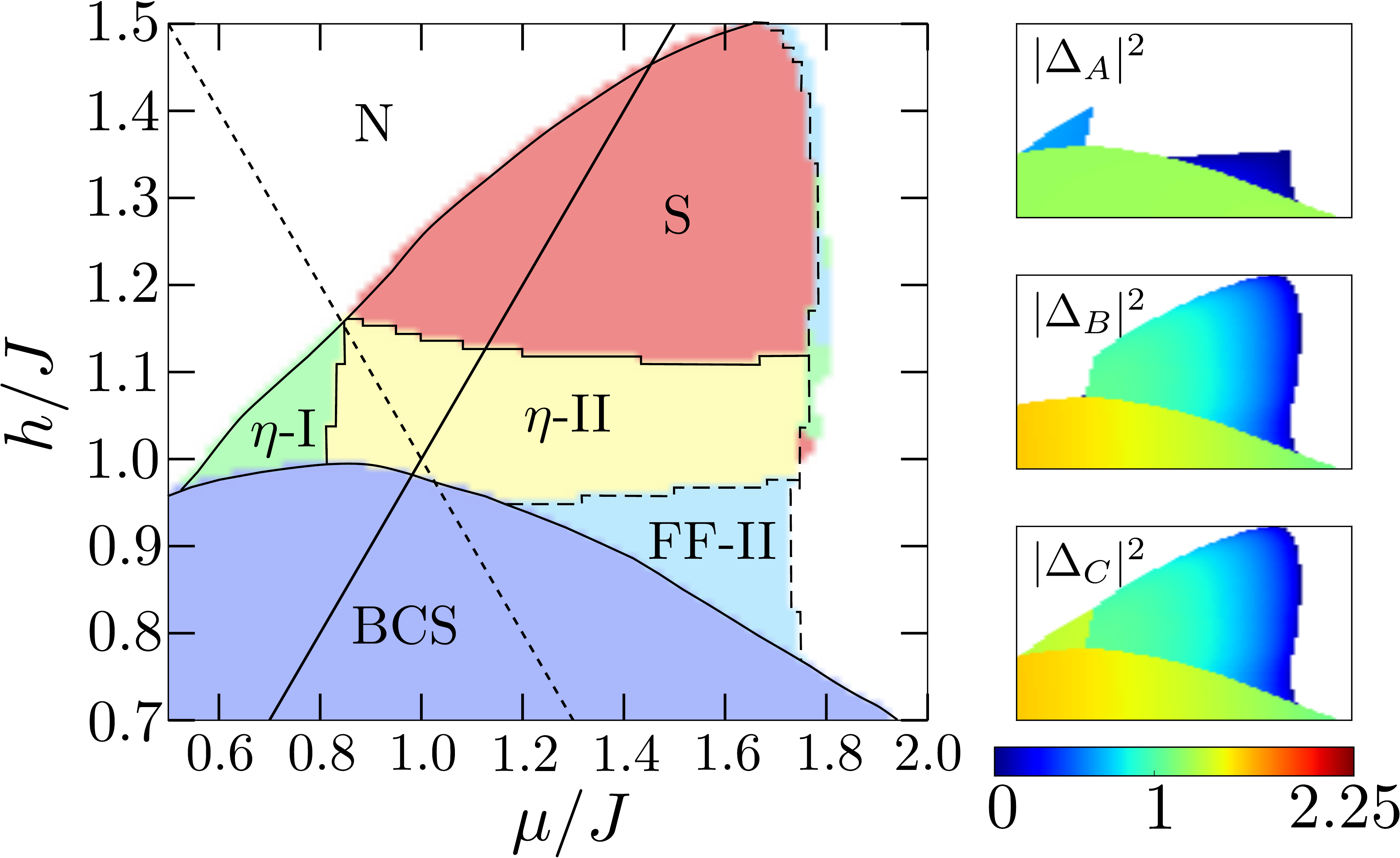}
\caption{Phase diagram at $U = -4 J$ and $\beta J = 10$ around the intersection of the Van Hove and the FB singularity lines. The area around this intersection features enhanced superfluidity and a region with nontrivial superfluid phases, wherein we can identify the following phases: FF-II with the FF momentum $\q$ directed along the diagonal of the unit cell, $\eta$-II where the momentum $\q$ saturates at the $M = (\pi, \pi)$ point of the Brillouin zone, $\eta$-I where $\q$ saturates at the $X = (\pi, 0)$ point, and a Sarma phase (S), where the spin imbalance is present but no relative shift of the two FSs. Solid and dashed lines mark first order and continuous phase transitions respectively. In the panels on the right hand side, the value of the three components of the order parameter, $\boldsymbol{\Delta}$, are plotted. In the $\eta$-II and Sarma phases $\Delta_A = 0$, while in the $\eta$-I phase $\Delta_B = 0$. The order parameters close to the steep right-hand side boundary of the imbalanced superfluidity lobe are small and go to zero as we approach the continuous phase transition to the normal phase; therefore, at this boundary, there is visible numerical noise.}
\label{fig.phd}
\end{figure}
\begin{figure}
\includegraphics[width = \columnwidth]{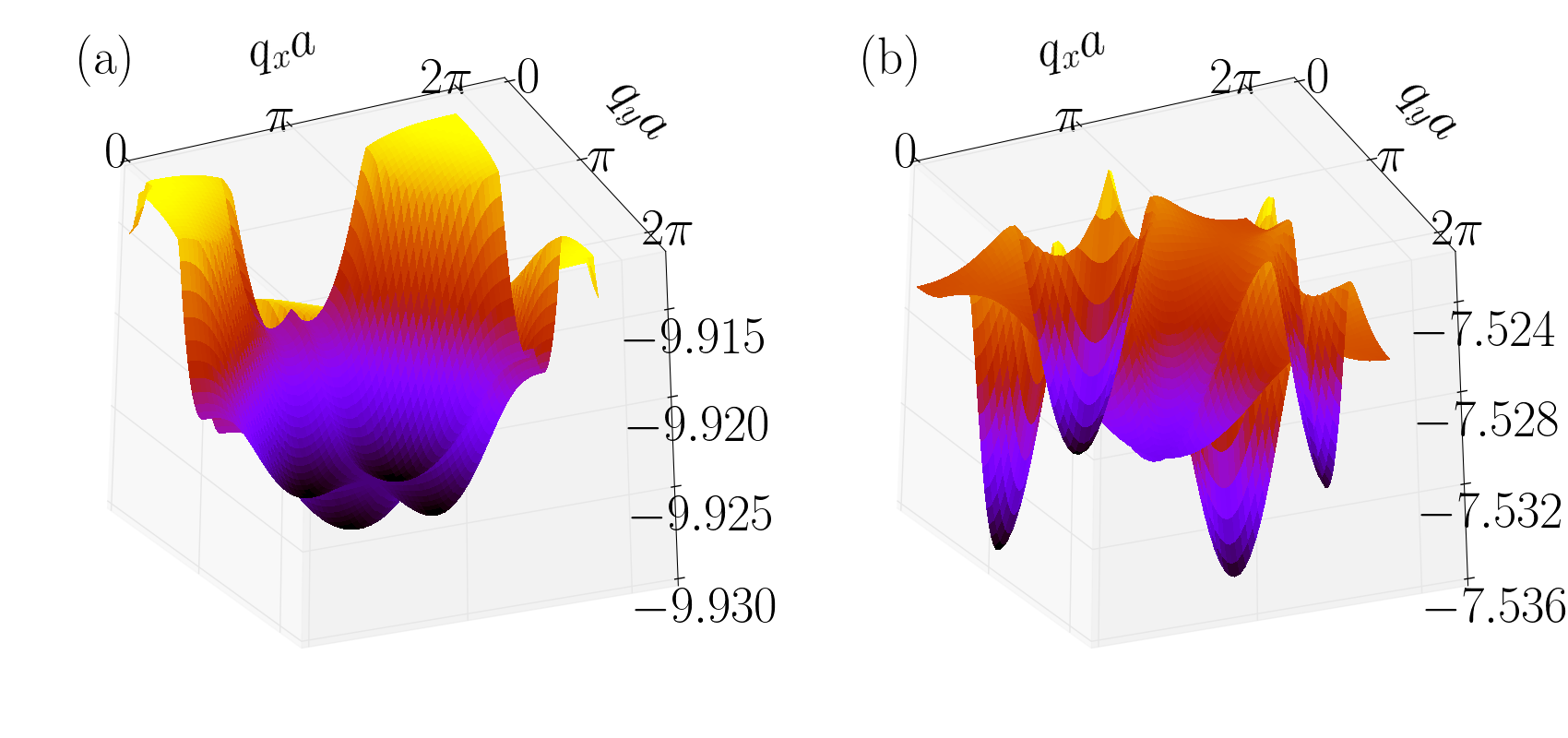}
\caption{Thermodynamic potential, $\Omega(\q)$, as a function of the FF momentum $\q$ at (a) $\mu = 1.4\, J$ and $h = 0.9\, J$, which is minimized by the FF-II phase and at (b) $\mu = 0.8\, J$ and $h = 1.05\, J$, which is minimized by the $\eta$-I phase. }
\label{fig.omega-eta}
\end{figure}
\begin{figure*}
\includegraphics[width = .99\textwidth]{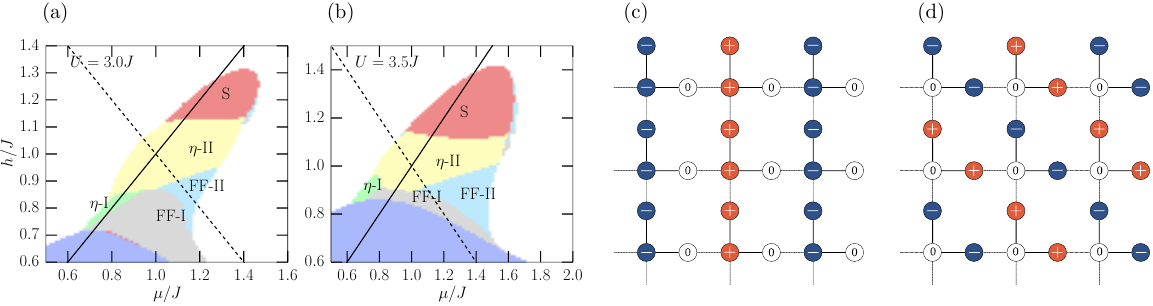}
\caption{Phase diagrams for interactions: (a) $U = -3\, J$ and (b) $U = -3.5\, J$. Unlike in Fig.~\ref{fig.phd}, there is visible FF-I phase. Also, with decreasing interaction strength, the region with imbalanced superfluidity shrinks until it disappears completely. In panel (c) and (d) we show the configurations the order parameter assumes in the $\eta$-I and $\eta$-II phases, respectively. }
\label{fig.otherphd}
\end{figure*}
\section{Phase diagrams}
\label{sec.phd}
We start with a non-dimerized lattice $\delta_{x, y} = 0$, and study the possible phases as a function of $\mu$ and $h$. To this end we minimize the thermodynamic potential Eq.~(\ref{eq.thermpot}) by both (1) solving the {\it gap equation}
\begin{equation}
\frac{\delta \Omega^{\phantom{\dag}}}{\delta \bf \Delta^\dag} = 0, \quad \frac{\delta \Omega}{\delta \bf q} = 0
\label{eq.gapeq}
\end{equation}
and (2) by simplex minimization with the {\it Nelder-Mead algorithm}, in order to make sure we find the global minimum in the multidimensional landscape of $\Omega(\bf \Delta, q)$; in principle, Eq.~(\ref{eq.gapeq}) identifies only a stationary point. 

The resulting phase diagram is shown in Fig~\ref{fig.phd}, for $U = -4\, J$. Since the band structure of the Lieb lattice is symmetric with respect to reflection by the $\varepsilon = 0$ plane, it has a particle-hole symmetry. This symmetry, together with the symmetry of renaming the spin species, manifests itself in a phase diagram as a symmetry with respect to transformation: $\mu \to -\mu$ and $h \to -h$. Therefore, it is sufficient to consider only one quarter of the phase diagram, here $\mu \geq 0$ and $h \geq 0$. We focus on the region around the crossing of the FB and Van Hove singularity lines, wherein we expect the nontrivial superfluid phases to be favoured. Indeed, apart from the BCS phase for low chemical potential imbalance, we find a region with a plethora of non-trivial superfluid phases. As already reported in~\cite{Huhtinen2018}, above the BCS region we find the Fulde-Ferrell phase, whose momentum $\bf q$ grows with a growing imbalance until it saturates at the edge of the Brillouin zone ($M$ point in this case), where the phase is called the $\eta$-phase. In the $\eta$ phase, the modulation of the order parameter is twice the periodicity of the lattice. In general, we find two types of the FF and $\eta$ phases which differ by the direction of the momentum $\q$. The FF and $\eta$ phases where $\q$ is directed along one of the lattice vectors (${\bf a}_x$ in our case) will be called FF-I and $\eta$-I phases, respectively. The other direction of $\q$ is along the diagonal of the unit cell ($({\bf a}_x + {\bf a}_y) / \sqrt2$), and respective phases will be called FF-II and $\eta$-II. In the phase diagram of Fig.~\ref{fig.phd} the FF-I phase is not present, but it can be found for lower interactions, see Fig.~\ref{fig.otherphd} (a) and (b) for $U = -3\, J$ and $U = -3.5\, J$. 

The mechanical stability of the phases requires the inverse compressibility matrix $\propto \partial \mu_i / \partial n_j$ to be positive definite. We checked numerically that this is indeed the case for the imbalanced superfluid phases (FF, $\eta$ and Sarma phase) from Fig.~\ref{fig.phd}; these phases are stable against phase separation. For the BCS phase one can present a simple analytical argument: within this phase the magnetization identically vanishes, $m(\mu_\up, \mu_\down) \equiv 0$, and therefore $\partial m / \partial \mu_\up = \partial m / \partial \mu_\down = 0$ and $(\boldsymbol{\kappa})_{ij} = \partial n_i / \partial \mu_j$ can be written as
\begin{equation*}
\boldsymbol{\kappa}^{\rm BCS} = \frac14 \begin{bmatrix}
(\partial_\mu + \partial_h) n & (\partial_\mu - \partial_h) n \\
(\partial_\mu + \partial_h) n & (\partial_\mu - \partial_h) n \\
\end{bmatrix} ~,
\end{equation*}
with $n$ being the total density, and $\partial_\mu = \partial / \partial \mu$, $\partial_h = \partial / \partial h$. The rank of the above matrix is clearly $rk\, \boldsymbol{\kappa} = 1$; therefore, within the BCS phase the eigenvalues of $\boldsymbol \kappa$ are $\frac12 \partial n / \partial \mu$ and zero, reflecting the incompressible (gapped) spin mode of the BCS state.

The area in the phase diagram of generic FF phases shrinks with increasing interaction (in magnitude), while the corresponding $\eta$ phases grow in size. Also, the relative area of the Sarma phase grows with interaction, as predicted by energy estimation in Ref.~\cite{Liu2003}. We specifically checked that these phases correspond to global minima of $\Omega(\q)$. For example, the $\eta$-I and $\eta$-II phases in Fig.~\ref{fig.phd} correspond to different global minima, and the phase transition between them is discontinuous. On the other hand, the transition between the FF-II phase and the $\eta$-II phase is continuous, and the vector $\q$ smoothly grows until it reaches its maximum value. The phase transitions from the BCS phase to the FF and $\eta$ phases are first order, as are the transitions to the normal phase, apart from the steep right-hand-side boundary of the imbalanced superfluidity lobe, where the transition is continuous. Figure~\ref{fig.omega-eta} shows the thermodynamic potential as a function of $\q$ for the FF-II phase, where it is minimized by $\q = (\pi \pm q_0, \pi \pm q_0)$ and for the $\eta$-I phase, minimized by $\q = (\pi, 0)$ or $(0, \pi)$.

For larger imbalances of chemical potentials, $h$, we find a stable Sarma phase, which is characterized by an imbalance in densities of the two spin components, but accomodated in such a way, that the order parameter remains spatially uniform, i.e. $\q = 0$. This is in contrast to other works, where the existence of a stable Sarma phase required fixed densities (canonical ensemble). The phase transition between the $\eta$-II and Sarma phases is discontinuous, as it involves an abrupt jump in the magnitude of $\q$. 

\begin{figure}
\includegraphics[width = \columnwidth]{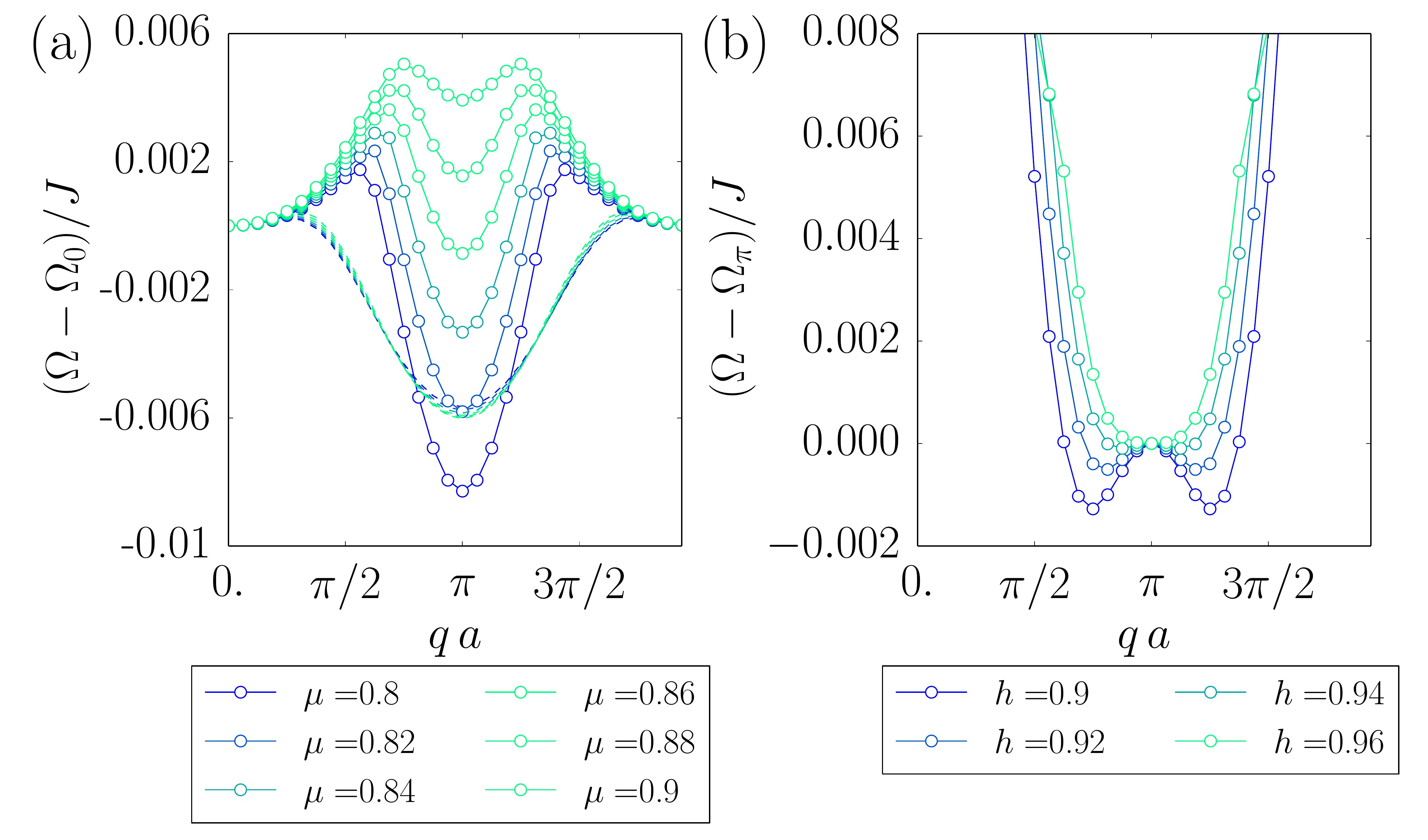}
\caption{(a) The thermodynamic potential as it changes across the $\eta$-I and $\eta$-II phases. At the transition point the minimum for the parallel q vector goes below that for the diagonal $\q$ vector. The $\eta$-I phase becomes stable and the transition is clearly discontinuous. (b) The thermodynamic potential across the FF and $\eta$-II phases as a function of $q = |\q|$. The FF wave vector for which the minimum is attained converges continuously to $\pi$, and the double-well shape of the thermodynamic potential suggests a Landau mechanism of a continuous phase transition. }
\label{fig.phasetrans}
\end{figure}
\begin{figure}
\includegraphics[width = .99\columnwidth]{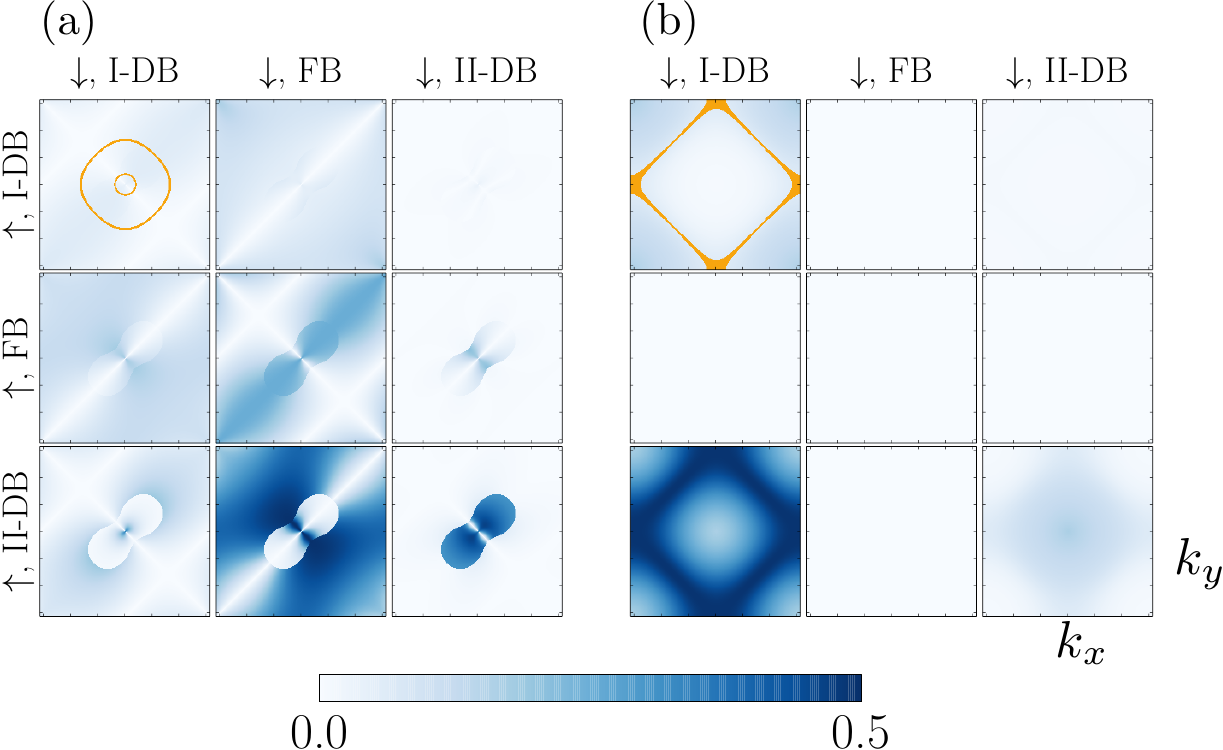}
\caption{Pairing between various bands for the $\eta$-II phase in panel (a) and the small $\eta$-II region for $\mu = 0$ and $h = 2\, J$ in panel (b), see text. In (a) we can separate two contributions to the pairing: an interband pairing between the FB and II-DB and an intraband pairing, mostly within the II-DB. There is also a visible trace of deformation of the majority FS. In (b), the pairing is between particles of opposite spins residing on two different dispersive bands: I-DB and II-DB. The two FSs are identical, as they are formed at $\mu_{\up \down} = \pm 2\, J$. They are far away from the FB, which does not contribute to the pairing in this case. }
\label{fig.correl}
\end{figure}
\section{Fulde-Ferrell and $\eta$ phases}
\label{sec.ff}
The FF-II phase in the phase diagram in Fig.~\ref{fig.phd} is characterized by a momentum $\q$ which is diagonal with respect to the unit cell of the lattice, that is $q_x = q_y \neq 0$. This momentum grows with increasing chemical potential imbalance $h$ until it reaches the $M$ point of the Brillouin zone and remains there when $h$ is further increased. This phase with momentum $\q$ saturated to its maximum value is the $\eta$-phase. 

Remarkably, we find two different types of FF phases: FF-I and FF-II, and two types of $\eta$ phases: $\eta$-I and $\eta$-II; for different values of chemical potentials different directions of the FF vector minimize the thermodynamic potential $\Omega(\q)$. In the $\eta$-I phase the centre-of-mass momentum of Cooper pairs, $\q$, is parallel to one of the lattice vectors, in this case ${\bf a}_x$ (which is a degenerate choice with one for $\q$ in the direction of ${\bf a}_y$). This phase occupies a relatively small corner of the phase diagram. As schematically shown in Fig.~\ref{fig.otherphd}(c), the order parameter at site B, i.e. the site neighbouring the central site A in the direction of the momentum $\q$ ($x$ direction), is zero. The order parameter for sites A and C changes sign every time we translate it by the lattice vector ${\bf a}_x$. Thus, the modulation of the order parameter has a period twice that of the lattice, see Fig.~\ref{fig.otherphd}(c). 

In the $\eta$-II phase, the FF momentum is in the direction of the diagonal of a unit cell; it extends between the $\Gamma$ and $M$ points. In this case, it is the A site that supports zero order parameter, $\Delta_A = 0$, and for sites B and C the order parameter is equal in magnitude, $|\Delta_B| = |\Delta_C|$. There is a degeneracy of sign of the order parameter at these sites. We may fix it to be the same within a given unit cell, and this sign will change as we translate by either of the lattice vectors, ${\bf a}_x$ or ${\bf a}_y$. Thus, the sign of the non-zero order parameter may assume a chessboard-like pattern, see Fig.~\ref{fig.otherphd}(d). 

In Fig.~\ref{fig.omega-eta} we explicitly show the thermodynamic potential as a function of $\q$ for two cases: the FF-II phase where the momentum $\q$ is diagonal with respect to the unit cell and the $\eta$-I phase, where $\q$ has only one non-vanishing component. Clearly, for those values of momentum, the quantity $\Omega(\q)$ attains its global minimum. The phase transition between the $\eta$-I and $\eta$-II phases is of the first order, as can be seen from Fig.~\ref{fig.phasetrans}(a); as we cross between $\eta$-I and $\eta$-II, the global minimum of the thermodynamic potential initially corresponding to $\q = (\pi, 0)$ is surpassed by the local minimum at $\q = (\pi, \pi)$, which then becomes the new global minimum. In addition, there is a jump of both a total density and a polarization, $n_\up - n_\down$, as we cross the phase boundary. The phase transition between the FF-II phase and $\eta$-II phase is, on the other hand, continuous. The FF momentum $\q$ grows continuously from having equal values in one of the four degenerate minima around $(\pi, \pi)$ until it reaches a critical point with a single minimum at the $M$ point, see Fig.~\ref{fig.phasetrans}(b). As expected, no jump in density or polarization is present. The FF-I phase, where the vector $\q$ is parallel to one of the lattice vectors, but remains inside the Brillouin zone, is present for smaller (in magnitude) values of interaction, see Fig.~\ref{fig.otherphd}(a) and (b), but vanishes already at $U = -4\, J$. 

The reason behind such a rich phase diagram of FF-type phases is the presence of a FB, where particle density can reorganize without energy cost and therefore contribute to pairing with greater flexibility. In the FF and $\eta$-II phases, the density distribution of the minority component in the FB mimics the distribution of the majority component in the II-DB, as was already reported in~\cite{Huhtinen2018}. This allows for an interband pairing, i.e. pairing between atoms residing in different bands, the FB and II-DB in our example. 

Intraband pairing between particles within the same dispersive band is also possible. The extra atoms, which do not contribute to the pairing, remain in a normal state, forming a sharp Fermi surface. The density of this normal gas is constant and equal to
\begin{equation}
n_\up({\bf k}) - n_\down({\bf k}) = 1 ~,
\end{equation}
which, after integration, reads $N_\up - N_\down = {\rm Vol} / 4 \pi^2$, in accordance with the Luttinger theorem~\cite{Kinnunen2017}, which states that the number of states enclosed by the FS (${\rm Vol}$) does not change when the interactions are present. Within the region delimited by this FS, there is the intraband pairing of atoms within the II-DB. This mechanism can be understood from the calculation of the band-resolved correlations, $C^{n, m}_{\bf k, \bf q} = \langle d_{n\, \bf k \up} d_{m\, \bf q - k \down} \rangle$, wherein we can discern between the two most significant contributions to the pairing: the interband pairing between atoms in the FB and II-DB and the intraband pairing within the II-DB, as plotted in Fig.~\ref{fig.correl}(a) for the $\eta$-II phase. 

In the case of the generic FF phases, there is a relative momentum shift of the two FSs. Together with the deformation of the FS of the majority species, especially in the region of the other FS, it enhances the pairing in a similar way to the nesting effect in the square lattice FFLO mechanism. In our case, it contributes to the intraband pairing in the II-DB. In the interacting system, this deformed FS can be seen in the distribution of the noninteractng gas formed by the excess particles, but it also leaves its imprint on the momentum distribution of the paired particles and on correlations. For plots of relevant correlations and densities, see Ref.~\cite{Huhtinen2018}. There, we also showed how the deformation of the majority FS augments as one approaches the FB singularity, where the FS loses its continuity (and the minority FS disappears). 

There is also a possibility of a pairing between two dispersive bands, I-DB and II-DB. This happens when both chemical potentials are away from the FB, which therefore does not contribute to the pairing. Around the crossing of the two Van Hove singularities, at $\mu_{\up \down} = \pm 2\, J$ (or $\mu = 0$ and $h = 2\, J$), there is a small region with the $\eta$-II phase, where the pairing comes virtually exclusively from the correlations between I-DB and II-DB, see Fig.~\ref{fig.correl}(b). Since the FS around the Van Hove singularity assumes a particular square shape, the momentum shift of $\q = (\pi, \pi)$ maximizes the overlap between the two densities, because the density of the minority component fills the inner region delimited by such FS and the majority component fills the outer region. 

\begin{figure}[!h]
\includegraphics[width = .99\columnwidth]{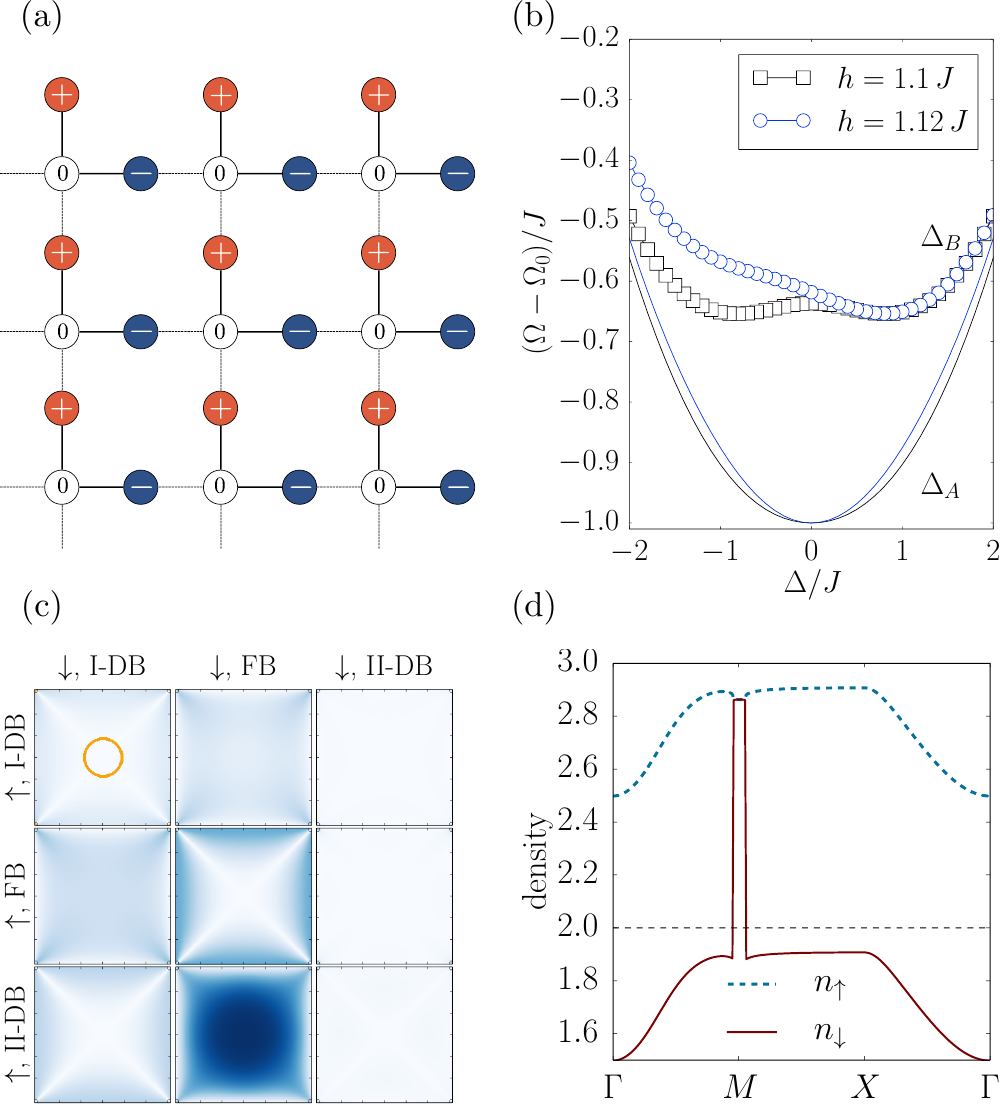}
\caption{(a) In the Sarma phase, the order parameter at the central site is zero, $\Delta_A = 0$, while it takes opposite but equal in magnitude values at sites B and C, $\Delta_B = -\Delta_C$. In panel (b) we show the dependence of $\Omega$ on the order parameters at fixed $\mu$ as the imbalance $h$ is being increased across the transition between the $\eta$-II phase and the Sarma phase. While the thermodynamic potential is always minimized by $\Delta_A = 0$, its symmetry with respect to $\Delta_B$ changes from a symmetric function to a single minimum function. In this way, the sign of $\Delta_B$ is determined as opposite to $\Delta_C$. This intracell modulation of the order parameter allows for the existence of the stable Sarma phase. In (c) we plot the band resolved correlations, $\langle d_{\alpha \bf k \up} d_{\beta \bf q - k \down} \rangle$, which show, that the main contribution comes from the pairing between majority atoms in the II-DB and minority atoms in the FB (colour scale as in Fig.~\ref{fig.correl}). (d) Cumulative densities for majority and minority components along the high symmetry lines: $\Gamma$--$M$--$X$--$\Gamma$. There is a clear matching of the density profiles where the interband pairing takes place, accompanied by a jump in densities $n_\up - n_\down = 1$, in accordance with the Luttinger theorem. Around the $M$ point (corners of the BZ) the densities are equal. }
\label{fig.sarma}
\end{figure}
\begin{figure}
\includegraphics[width = .99\columnwidth]{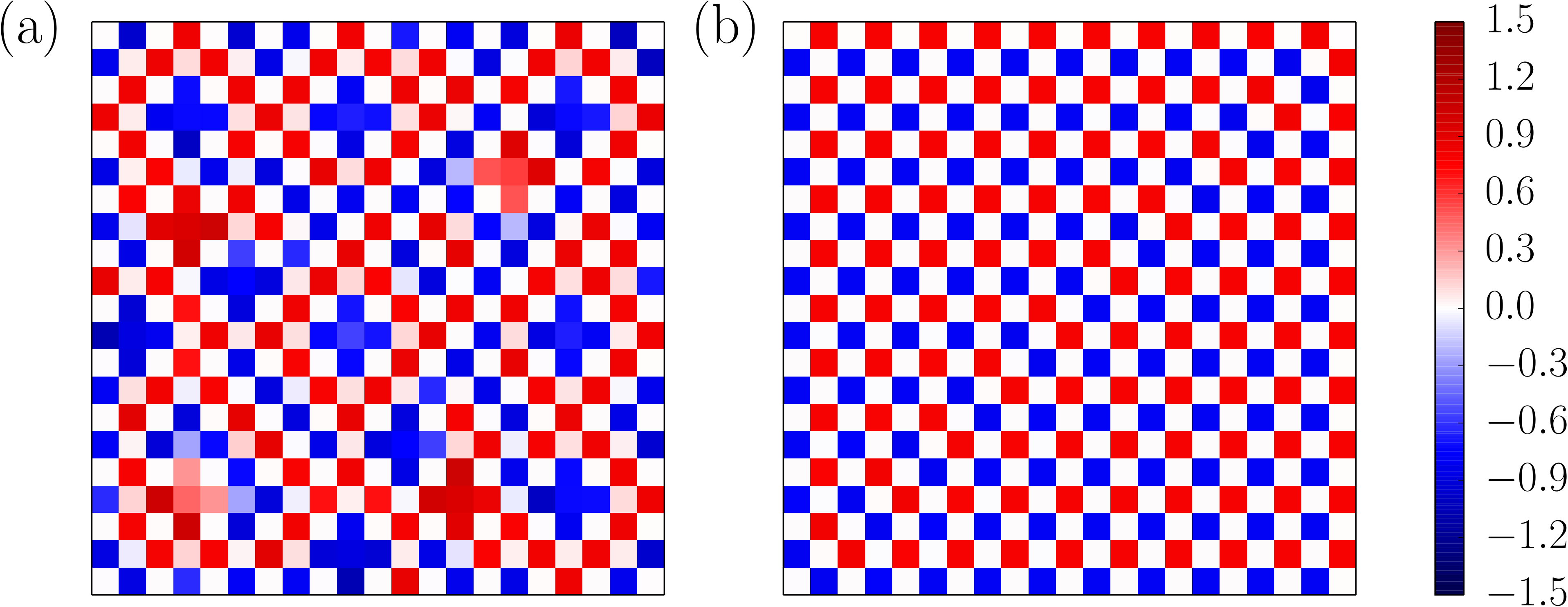}
\caption{Real space configuration of the order parameter $\Delta_{\bf i \alpha}$ obtained as a solution to Eq.~(\ref{eq.realbdg}). In (a) an eta-II phase is obtained for $\mu = 1.4\, J$ and $h = 1.05\, J$, and in (b) the solution converges to Sarma phase at $\mu = 1.4\, J$ and $h = 1.3\, J$. The shown central fragment comes from the simulation on a $32 \times 32$ lattice. }
\label{fig.real}
\end{figure}
\section{Sarma phase}
\label{sec.sarma}
When the chemical potential imbalance, $h$, is increased, $\mu_\down$ decreases and approaches the flat band, while $\mu_\up$ grows. For the non-interacting system it means a very low occupation of the II-DB for the minority component, which fills it only in small corners around the $M$ point, whereas the majority component fills most of the II-DB save a small region around the $\Gamma$ point. Thus, as we leave the $\eta$ phase by increasing the imbalance, it becomes energetically favourable for the system to switch to a pairing mechanism that does not involve either a momentum shift of the FSs or their deformation. Instead, almost all the pairing comes from the correlations between the FB and the II-DB. The correlations between particles in different bands are plotted in Fig.~\ref{fig.sarma}(c), while crossections of cumulative density profiles in Fig.~\ref{fig.sarma}(d). Like in the FF and $\eta$ phases, the pairing is manifested by matching density profiles, whereas the normal gas, which does not take part in the pairing, shifts the density of the majority component by $n_\up - n_\down = 1$ wherever the unpaired particles are present. The presence of a normal gas is also reflected in the gap closing in the quasiparticle energy spectrum around the $M$ point. This mechanism resembles the so-called interior gap pairing, known also as the Sarma phase. In its original form, it is a pairing between the atoms in the vicinity of the minority FS and atoms of opposite momenta from inside of the majority Fermi sea. In our case, it is the pairing between the FB and the dispersive band, but like in the original works by Sarma~\cite{Sarma1963} and by Liu and Wilczek~\cite{Liu2003}, no momentum shift is involved. 

An important question in relation to this phase is its stability. Fig.~\ref{fig.sarma}(b) shows the thermodynamic potential as a function of order parameters $\Delta_A$ and $\Delta_B$ as we cross from the $\eta$ phase to the Sarma phase. The global minimum of the thermodynamic potential yields a non-zero order parameter and a finite spin imbalance. The initial degeneracy of sign for $\Delta_B$ an $\Delta_C$ in the $\eta$ phase (we had $\Delta_B = \pm \Delta_C$) is replaced by a configuration, where $\Delta_B$ and $\Delta_C$ have opposite signs. The central site of the unit cell assumes $\Delta_A = 0$, like in the $\eta$-II phase. This explains the existence and stability of the Sarma phase in our multiband model: due to a unit cell composed of more than one site, an {\it intracell} modulation of the order parameter is possible, leading to a stable spin-imbalanced superfluid phase which does not involve a finite FF momentum (modulation across different unit cells). It is thus the non-uniformity within the unit cell that allows for the Sarma phase, see Fig.~\ref{fig.sarma}(a). 

Finally, we confirm the existence of the previously mentioned phases by converging the set of equations~(\ref{eq.realbdg}) on a $32 \times 32$ lattice. In Fig.~\ref{fig.real} the order parameters for the $\eta$-II phase and the Sarma phase are shown. In case of the Sarma phase, there is a clear alteration of the sign of the order parameter at sites B and C, while site A supports $\Delta_A = 0$. In so far as the $\eta$ phase is concerned, the order parameter's pattern also conforms to our predictions, with some defects being evidently visible, however. This might be a result of higher degeneracy of the sign choice in the $\eta$ phase and of the open boundary conditions we employed. It is important to note that the real-space computation may favour a different type of FFLO ansatz, such as the LO ansatz. In fact, we observe a spatial density modulation (not shown here), which accompanies the variation of the order parameter. 
\section{Temperature dependence}
\label{sec.temp}
\begin{figure}
\includegraphics[width = \columnwidth]{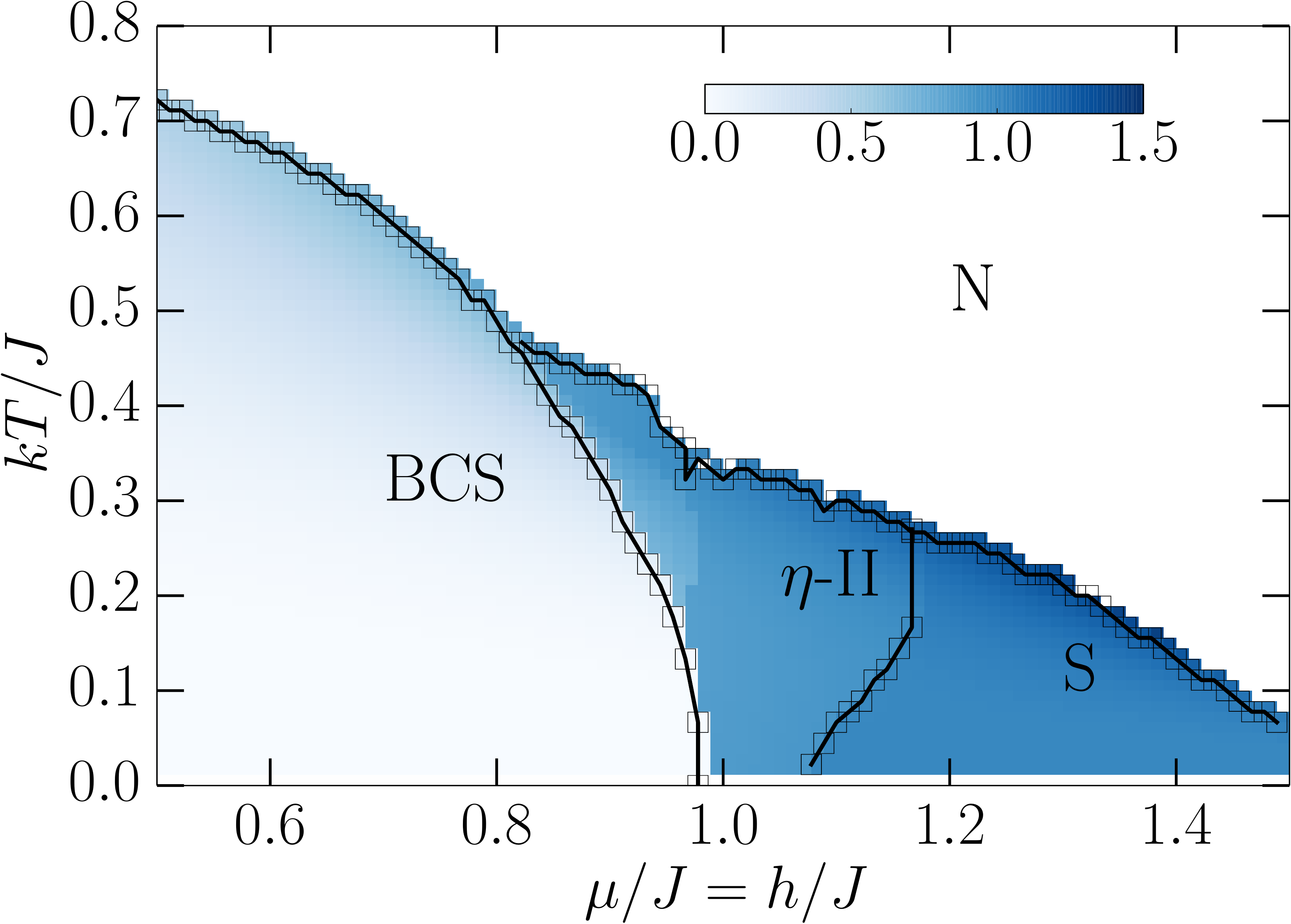}
\caption{Temperature dependence of the phase diagram at interaction $U = -4\, J$. The phase diagram is plotted against the temperature $kT / J$ and chemical potential $\mu / J$ along the FB singularity line $\mu = h$. Colour intensity denotes the polarization (apart from the normal phase), and lines show the detected phase boundaries.}
\label{fig.temp}
\end{figure}
Next, we turn to the temperature dependence of the phase diagram. In Fig.~\ref{fig.temp} we show the phase diagram plotted against the temperature and chemical potential as we move along the FB singularity line, $\mu = h$. This choice has been made because the superfluidity enhancement and non-trivial superfluid phases with spin imbalance occur in the vicinity of this line. 

It is clear that with growing temperature the area where we may observe non-trivial superfluid phases, such as FF, $\eta$ or Sarma phase, shrinks. This region disappears already at lower temperatures than the BCS lobe. At higher temperatures, where the BCS lobe already occupies a small region in the $\mu$ -- $h$ phase diagram, there is visible spin imbalance (shown as colour intensity in Fig.~\ref{fig.temp}), for at high temperatures single-particle excitations are allowed. The Berezinskii-Kosterlitz-Thouless (BKT) temperature for superfluids in two dimensions is smaller than the BCS one, but can also be of the same order of magnitude~\cite{Peotta2015,Julku2016,Liang2017,Julku2018}.

\section{Staggered hoppings}
\label{sec.stag}
\begin{figure}
\includegraphics[width = .99\columnwidth]{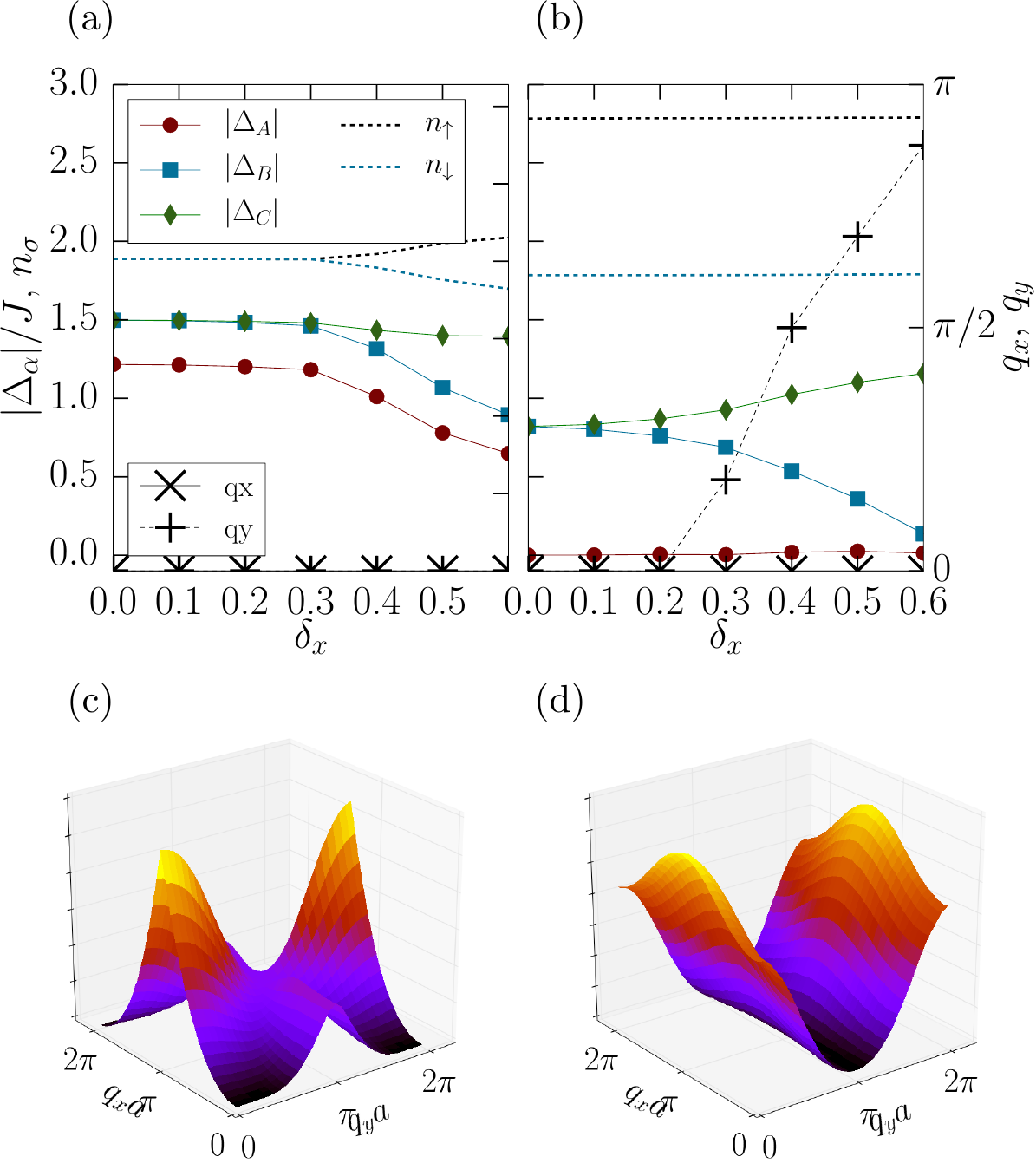}\\
\caption{Order parameters (solid lines with markers), densities (dashed lines) and FF momentum components (dashed lines with markers) of (a) the BCS phase ($\mu = 0.8\, J$ and $h = 0.8\, J$) and (b) the Sarma phase ($\mu = 1.4\, J$ and $h = 1.3\, J$), as the staggering parameter $\delta_x$ is increased, while $\delta_y = 0$. The nature of the phase can change with growing staggering of the hoppings, as detailed in the text. In panels (c) and (d): thermodynamic potential, $\Omega(\q)$, as a function of $\q$ for a dimerized lattice at $\mu = 1.4\, J$ and $h = 1.3\, J$, corresponding to the Sarma phase. With growing dimerization in the $x$-direction ($\delta_y = 0$), the initial configuration favouring the Sarma phase at $\delta_x = 0.3$ (c) turns into one whose minimum does not depend on $q_x$, here $\delta_x = 0.7$ (d). }
\label{fig.delx}
\end{figure}
Finally, we reintroduce the lattice dimerization parameters $\delta_{x, y}$. First, we consider staggering only in one direction: $\delta_x \neq 0$, while $\delta_y = 0$. In this case, with growing asymmetry, the two directions are no longer eqivalent, and this takes effect on the FF momentum, $\q$. In Fig.~\ref{fig.delx}(a) we show what happens to the BCS solution at $\mu = 0.8\, J$ and $h = 0.8\, J$ when the asymmetry is switched on and increased. Since the B and C sites are not equivalent anymore, there is no reason to expect the order parameters $|\Delta_B|$ and $|\Delta_C|$ to be equal. Indeed, with growing $\delta_x$ the initial degeneracy is removed. This is associated with a density imbalance; therefore, as a consequence of growing asymmetry, the system enters a Sarma phase for fixed values of chemical potentials. 

The effect of hopping asymmetry for the Sarma phase, initially present at $\mu = 1.4\, J$ and $h = 1.3\, J$, is shown in Fig.~\ref{fig.delx}(b). Here, the effect is even more dramatic: not only the magnitudes of the order parameters start diverting from one another, but at a certain $\delta_x$ a non-zero FF momentum appears, growing as the staggering parameter is increased. It is not surprising that it is the $q_y$ component that assumes a non-zero value, since the system becomes weakly coupled in the $x$-direction when the staggering in on, which effectively, gradually reduces its dimesion to one. As seen in Fig.~\ref{fig.delx}(c) and (d) the thermodynamic potential, $\Omega(\q)$, becomes independent of $q_y$ for large enough values of $\delta_x$. At $\delta_x \approx 0.7$ (see Fig.~\ref{fig.delx}(d)) the momentum $q_y$ reaches the value $q_y = \pi$ and the sytem enters the $\eta$-I phase. Finally, at values of the staggering close to $\delta_x \approx 1$ we have a collection of independent one-dimensional wires. 

When both staggering parameters are set to non-zero values, the scenario is similar save the fact, that no direction for $\q$ is now preferred. However, since now as the system becomes decoupled, it approaches a collection of disconnected unit cells, for which the mean-field theory loses its sense, and the system should be rather described at a level of single-particle physics. 

\section{Conclusions and prospects}
\label{sec.conclusions}
In conclusion, we studied superfluid phases of the Hubbard model with spin imbalance on a Lieb lattice. We found, that if one of the chemical potentials is in the vicinity of a FB, a spin-imbalanced superfluidity is possible. By considering a mean-field BCS theory and a Fulde-Ferrell ansatz for the order parameter, we found four types of phases with spatial modulation of the order parameter and a Sarma phase, where the density imbalance does not require a spatial modulation. By studying spin resolved correlations, we discover, that the mechanism responsible for these effecs involves pairing between different particles in different bands of the system: a FB, where the atoms can readjust their momentum space density with no energy cost, and dispersive bands. On the one hand, such a flexibility allows for the observed variety of the imbalanced superfluid phases, especially the FF and $\eta$ phases with spatial modulation of the order parameter. On the other hand, the multiorbital structure of the unit cell allows for the modulation of the order parameter within the unit cell, which in turn stabilizes the Sarma phase. We also considered the effects of finite temperature on the discovered phases. Finally, we assumed the staggered form of hopping between the neighbouring sites, and we concluded that such dimerized hopping can change the phase of the system. 

In summary, the Lieb lattice offers an astonishing richness already at the mean-field level, and certainly deserves further study, especially that ultracold Fermi gases offer the possibility of immediate realization of our predictions. Lieb lattice geometry has been realized in optical lattices~\cite{Taiee1500854,PhysRevLett.118.175301} and novel techniques such as digital mirror devices or holograms~\cite{Hueck2017,Gauthier2016,Zupancic2016} permit further experiments with lattices with complex unit cells. 

\bigskip
{\bf Acknowledgments ---} We are grateful to K.-E. Huhtinen for reading the manuscript. This work was supported by the Academy of Finland through its Centres of Excellence Programme (2012–-2017) and under project NOs. 284621, 303351 and 307419, and by the European Research Council (ERC-2013-AdG-340748-CODE). Computing resources were provided by the Triton cluster at Aalto University.

\end{document}